# Scaling near the upper critical dimensionality in the localization theory

I. M. Suslov*)

*P. L. Kapitza Institute for Physical Problems, Russian Academy of Sciences, 117334 Moscow, Russia*
(Submitted 21 October 1997)
Zh. Éksp. Teor. Fiz. **113**, 1460–1473 (April 1998)

The phenomenon of upper critical dimensionality $d_{c2}$ has been studied from the viewpoint of the scaling concepts. The Thouless number $g(L)$ is not the only essential variable in scale transformations, because there is the second essential parameter connected with the off-diagonal disorder. The investigation of the resulting two-parameter scaling has revealed two scenarios, and switching from one to another scenario determines the upper critical dimensionality. The first scenario corresponds to the conventional one-parameter scaling and is characterized by the parameter $g(L)$ invariant under scale transformations when the system is at the critical point. In the second scenario, the Thouless number $g(L)$ grows at the critical point as $L^{d-d_{c2}}$, which leads to a violation of the Wegner relation $s=\nu(d-2)$ between the critical exponents for conductivity $s$ and localization radius $\nu$, which takes the form $s=\nu(d_{c2}-2)$. The resulting formulas for $g(L)$ are in agreement with the symmetry theory suggested in a previous publication, I. M. Suslov, Zh. Éksp. Teor. Fiz. **108**, 1686 (1995) [JETP **81**, 925 (1995)]. A more rigorous version of Mott's argument concerning localization due to topological disorder has been proposed. © *1998 American Institute of Physics.* [S1063-7761(98)02404-4]

## 1. INTRODUCTION

The one-parameter scaling hypothesis[1] has played an important role in development of the contemporary localization theory[2–8] and stimulated creation of the theory of quantum corrections[9] unambiguously supported by an experiment. The criticism of the one-parameter scaling[10–13] in fact refers not to underlying physical ideas, but rather to its justification in the formalism of $\sigma$-models.[14–16] The justification problem remains a pressing one, and may require more accurate definitions of the basic notions as well as lead to a restriction of the range of applicability. Here we discuss modifications of scaling concepts that we believe are inevitable in high-dimensional spaces.

Experience with phase-transition theory[17,18] indicates that scaling is applicable only to spaces with dimensionalities $d$ within an interval between the upper and lower critical dimensionalities, $d_{c1}$ and $d_{c2}$. For $d<d_{c1}$, there is no phase transition, and for $d>d_{c2}$, the mean-field theory is valid. There is no doubt that $d_{c1}=2$ in the localization theory,[1] whereas the issue of the upper critical dimensionality has remained a subject for discussions for many years.[19–25] As concerns the problem of the density of states (determined by the averaged Green's function $\langle G \rangle$), a comprehensive solution was recently found[26–29] by the author of this paper. It was demonstrated that $d_{c2}=4$ and how the condition $d>4$ simplifies the problem. The singularity at $d=4$ was also investigated, and the $(4-\epsilon)$-dimensional theory was developed. As concerns conductivity, which is determined by correlator $\langle G^R G^A \rangle$, the upper critical dimensionality could be, in principle, different for this quantity. The latter statement was made in Ref. 21, but there are some serious errors.[26] In fact, this conjecture is not true: the special role of dimensionality $d=4$ is a fundamental fact manifesting itself in the renormalizability of the theory,[26–29] and the renormalization properties of both density-of-states and conductivity problems are similar. This clearly follows from the fact that the same diagrammatic technique is used in both problems. Non-renormalizability of the theory at $d>4$ indicates the importance of the Hamiltonian structure on the atomic scale, which is the reason why the scaling invariance is broken. This reasoning is supported by the previously developed "symmetry theory,"[30] which yields the results that are in agreement with those of a one-parameter scaling only for $d<4$.

The present paper was motivated by two factors. On one hand, the opinion that $d_{c2}=\infty$ has recently become quite popular.[10,24,25] This opinion is not absolutely groundless since the one-parameter scaling theory *gives no indication* of the existence of an upper critical dimensionality. So there are certain drawbacks in the existing physical picture of localization, although it remains unchanged after many years of discussions.

On the other hand, the Wegner relation

$$s=(d-2)\nu \tag{1}$$

between the critical exponents for conductivity ($s$) to those of localization radius ($\nu$), which derives from the scaling theory, can be obtained under less demanding conditions.[31] Namely, it suffices to postulate the symmetry of correlation length on both sides of the transition and independence of the Thouless number at the critical point of the length scale. These two assumptions are taken for granted, so the mechanism responsible for a violation of the Wegner relation at $d>4$[23,30,32] deserves a consideration on the physical level.

The aim of the reported work was to fill these gaps and investigate the phenomenon of the upper critical dimensionality from the standpoint of the scaling concepts.

## 2. PROBLEM OF THE SECOND PARAMETER

The scaling theory[1] is based on the Thouless scaling consideration,[4,33] which is similar to the well-known Kadanov scheme in the theory of critical phenomena.[17,18] The disordered system in question, which is described by the Anderson model on a $d$-dimensional cubic lattice with the coupling integrals $J$ between nearest neighbors and the spread of the energy levels $W$, is divided into blocks of size $L$. In the absence of interaction between the blocks, the system has random energy levels with a characteristic spacing $\Delta(L) \sim J(a_0/L)^d$, where $a_0$ is the lattice constant. If the interaction is "switched on," the matrix elements between the states of the neighboring blocks appear and result in hybridization of "block" functions. The hybridization is the strongest between the states with close energies, and on a qualitative level we can consider only such states. By selecting in each block a level closest to a given energy $E$, we obtain the effective Anderson model with the spread of levels $W(L) \sim \Delta(L)$ and coupling integrals $J(L)$ determined by the corresponding matrix elements. The effective Anderson model provides a reduced description of the system on scales larger than $L$, and its properties are controlled by the Thouless number

$$g(L) = \frac{J(L)}{W(L)}, \tag{2}$$

related to the conductance $G(L)$ of a block with dimension $L$:

$$g(L) \sim \frac{\hbar}{e^2} G(L), \quad G(L) = \sigma(L) L^{d-2}. \tag{3}$$

Repeating the Thouless consideration for the effective Anderson model, we obtain an algorithm for calculating $g(bL)$ with integer $b$, given $g(L)$:

$$g(bL) = F(b, g(L)). \tag{4}$$

Abrahams *et al.*[1] considered the limit $b \to 1$ for this equation, when it can be rewritten in the form suggested by Gell-Mann and Low:

$$\frac{d \ln g}{d \ln L} = \beta(g). \tag{5}$$

The transition point $g_c$ is determined by a condition $\beta(g_c) = 0$, and the conductivity $\sigma = \lim_{L \to \infty} \sigma(L)$ and localization radius $\xi$ behave in the vicinity of the transition as

$$\sigma \propto (g_0 - g_c)^s, \quad \xi \propto (g_c - g_0)^{-\nu}, \tag{6}$$

where $g_0$ is the value of $g(L)$ at $L \sim a_0$, $1/\nu = g_c \beta'(g_c)$, and the critical exponent $s$ is determined by Eq. (1).

The theory developed by Abrahams *et al.*[1] corresponds to the simplest scenario of one-parameter scaling. In principle, one can imagine alternative situations. For example, if *two parameters*, $g(L)$ and $h(L)$, are important, we have, by analogy with Eq. (4),

$$g(bL) = F(b, g(L), h(L)), \quad h(bL) = G(b, g(L), h(L)), \tag{7}$$

which in the limit $b \to 1$ yields

$$\frac{d \ln g}{d \ln L} = \beta(g, h), \tag{8a}$$

$$\frac{d \ln h}{d \ln L} = \gamma(g, h), \tag{8b}$$

and the results are determined by the properties of two functions, $\beta(g, h)$ and $\gamma(g, h)$.

The arguments presented in Ref. 1 in favor of *one parameter* $g(L)$ scaling in spite of their peculiarity[1)] were well grounded. If the basic physical concept proposed in that paper is correct, the parameter $g(L)$ changes over a distance of the order of the correlation radius $\xi$, which can be arbitrarily large near the transition point, and Eq. (8b) can be analyzed at a constant $g$. If parameter $h(L)$ varies between the finite limits and is a monotonic function, it should on a certain scale $L_0 \ll \xi$ tend to a limiting value $h_\infty(g)$, and after substituting this value into Eq. (8a) we return to a one-parameter scaling. An oscillating behavior of parameter $h(L)$ would only indicate its inadequate definition, since averaging out the oscillations[34] would lead to an equation system like (8) with a smoothed parameter $\bar{h}(L)$, which varies monotonically. The parameter $h(L)$ can only be important if it tends to zero or infinity, but then can be detected on the level of order-of-magnitude estimates, and it would have had a clear physical sense. The entire scientific community has failed to suggest such a parameter throughout the period starting with the year 1979.

There are two candidates to the role of the second parameter which appear as a matter of course, but are rejected after a closer scrutiny.

a) While the Thouless scheme is constructed without approximations, the effective Anderson model contains a large number $n(L)$ of levels at each lattice site, which increases with $L$ and can be considered as the second parameter. But hybridization of states in neighboring blocks with energies $E$ and $E'$ is determined by the parameter $J(L)/|E - E'|$ and is inessential for $|E - E'| \gg J(L)$. Therefore one can take into account only $n(L) \sim J(L)/\Delta(L)$ levels around energy $E$, and the parameter $n(L)$ does not generate a new scale since it is of the same order as the Thouless number $g(L)$. Nonetheless, this modification of the Thouless scheme reveals new opportunities and will be considered in future work.

b) The overlap integrals in the Thouless construction are random values, and the ratio $\varphi(L) = \delta J(L)/J(L)$ between their fluctuation $\delta J(L)$ and their typical value $J(L)$ can be treated as the second parameter. But fluctuations can be neglected if $\delta J(L) \ll J(L)$, and the opposite case $\delta J(L) \gg J(L)$ is impossible since the extreme limit of off-diagonal disorder corresponds to a symmetric distribution of coupling integrals around zero when $\delta J(L) \sim J(L)$. Hence, the parameter $\varphi(L)$ can only play some role when it is of the order of unity and does not generate a new scale. Nonetheless, the off-diagonal disorder is significant, although a more appropriate definition of the corresponding parameter is required.

Estimates based on the optimal fluctuation technique[35,36] show that a typical wave function of localized states has a behavior

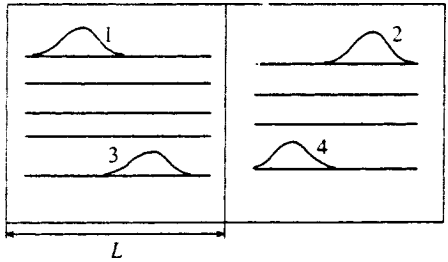

FIG. 1. At large $\zeta$ in Eq. (10), the block eigenfunctions are highly localized on scales $L<\xi$, which leads to strong off-diagonal disorder. For example, the overlap between the states 1 and 2 is substantially smaller than between 3 and 4.

$$|\Psi(r)|\propto\begin{cases}r^{-\zeta}, & r\ll\xi\\ \exp(-r/\xi), & r\gg\xi,\end{cases} \quad (9)$$

where $\zeta=d-2$ increases with the space dimensionality. This result is valid outside the close neighborhood of the transition point, i.e., in the region similar to that where the Landau theory[7] can be used, but such results have a tendency to become rigorous in spaces with a high dimensionality. In the critical region, a similar result is associated with investigations of multifractal properties of the wave functions[37]

$$\langle|\Psi(r)|^2|\Psi(r')|^2\rangle\propto|r-r'|^{-\eta}, \quad |r-r'|\ll\xi, \quad (10)$$

where $\eta\sim\epsilon$ for $d=2+\epsilon$ and $\eta\sim 1$ for $d=3$, i.e., it also increases with the space dimensionality. Therefore, let us assume that Eq. (10) holds in the critical region and $\zeta$ increases without bound as $d$ increases, and let us consider whether this property can lead to a catastrophe. A large value of $\zeta$ means that the block wave functions in the Thouless scheme are strongly localized on a scale smaller than $\xi$ (Fig. 1), which leads to strong off-diagonal disorder. For example, the overlap integral coupling states 1 and 2 is much smaller than that coupling states 3 and 4. The anticipated catastrophe is a localization due to the pure off-diagonal disorder, which can occur even if $W(L)=0$, i.e., when the spread of energy levels is neglected. The Thouless number $g(L)$ in this case is infinite and cannot play any role, and the hybridization of block states is controlled by a different parameter related to off-diagonal disorder.

### 3. LOCALIZATION IN THE CASE OF OFF-DIAGONAL DISORDER

A possibility of localization due to off-diagonal disorder was discussed in connection with the problem of formation of an impurity band in a semiconductor, which in fact stimulated the creation of the localization theory.[38] An isolated impurity in a semiconductor can generate a state with energy $E_0$ within the band gap. When the concentration of such impurities is finite, they form an impurity band, which is described in the site representation by the Anderson model with off-diagonal disorder (sometimes this is termed the Lifshitz model[5]):

$$\sum_{n'}J_{nn'}\Psi_{n'}+E_0\Psi_n=E\Psi_n. \quad (11)$$

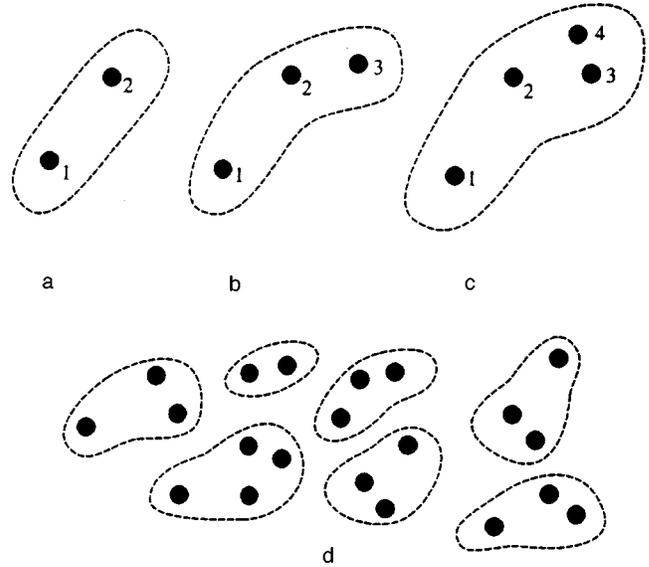

FIG. 2. Decomposition of an arbitrary configuration of impurities into clusters.

If the coupling integral drops exponentially,

$$J_{nn'}\propto\exp\{-\kappa|\mathbf{r}_n-\mathbf{r}_{n'}|\}, \quad (12)$$

where $\mathbf{r}_n$ is the coordinate of the $n$th impurity, the impurity band is completely localized in the limit of low concentration. Intuitive arguments in favor of this conjecture were suggested by Mott[3] on the basis of Lifshitz's classification of states.[35,36] Here we present a refined version of Mott's argument with the aim of attracting attention to physically significant aspects ignored by both Mott and Lifshitz.

The density of states $\nu(E)$ of the impurity band is a continuous function of energy and is formed by levels of which the overwhelming majority have energies different from that of an isolated impurity $E_0$. In order to obtain such levels, one should take into account the interaction between an arbitrary impurity atom 1 and its environment, no matter how weak it is. According to Lifshitz, the main factor is "collisions" between impurity atoms, i.e., random encounters among the latter. If the unit distance is the average distance between impurities, the limit of zero concentration corresponds to $\kappa\to\infty$ in Eq. (12). Since the overlap integral decays exponentially with the distance, only interaction between the nearest neighbors should be taken into account. Nevertheless, the analysis cannot be limited to pairwise "collisions."

Indeed, suppose that the nearest neighbor of atom 1 is atom 2. If the nearest neighbor of atom 2 is atom 1, the 1–2 pair can be treated in isolation from its environment (Fig. 2a). If the nearest neighbor of atom 2 is atom 3, we must consider the 1–2–3 cluster (Fig. 2b): first the hybridization of states of atoms 2 and 3 should be taken into account, then their interaction with atom 1. If the nearest neighbor of atom 3 is not atom 2 but atom 4, we must consider the 1–2–3–4 cluster (Fig. 2c), etc. If this construction process starts with atom 1 and ends with atom $i$, we consider by definition that atom 1 belongs to $i$th cluster. It is evident that atoms 2, 3, ... specified in this process belong to the same $i$th cluster.

Taking each impurity atom in turn as atom 1, we obtain a decomposition of an arbitrary configuration into clusters (Fig. 2d). The decomposition is unique since each atom in this scheme belongs to a certain cluster, and no atom can belong to two clusters at once (neglecting an infinitesimal probability of detecting an exact equality between two interatomic distances). Formally we should take into account arbitrarily large clusters, but in fact all clusters contain with an overwhelming probability a number of atoms on the order of unity (the existence of an infinite cluster would mean concentration of an infinite number of impurities in a finite volume).

Let us introduce parameter $R_1$, which is the characteristic interatomic distance inside a cluster, and parameter $R_2$, which is the characteristic separation between clusters. A thorough investigation is needed to give rigorous definitions of these parameters, but for any reasonable definition we have

$$R_1 < R_2, \qquad (13)$$

since clusters are formed from the nearest atoms.

By neglecting interaction between clusters and diagonalizing Hamiltonians of isolated clusters, we obtain the zero approximation for the density of states $\nu(E)$ of the impurity band, whose width is determined by the parameter $\exp(-\kappa R_1)$. This approximation is asymptotically exact in the limit of zero concentration, since the nearest neighbor of each atom is in the same cluster, and the shift of its level with respect to $E_0$ is calculated correctly in the lowest approximation.

Regarding each cluster as a site of a new lattice and taking into account interaction between clusters, we obtain the effective Anderson model with the spread of levels $W \propto \exp(-\kappa R_1)$ and overlap integrals $J \propto \exp(-\kappa R_2)$. By virtue of Eq. (13), we have $J/W \to 0$ as $\kappa \to \infty$, and in the zero-concentration limit, all states are localized inside the clusters. The latter clarifies the physical sense of these clusters.

Thus, we have proved the basic feasibility of localization of all states due to the pure off-diagonal disorder. Note that the pattern of hybridization between the eigenstates of separate blocks (Fig. 1), neglecting the spread of energy levels and in the limit $\zeta \to \infty$, is similar to the case of topological disorder in a system of impurities with exponential overlap.

## 4. TWO-PARAMETER SCALING

In the presence of off-diagonal disorder, a disordered system can be characterized by two parameters:

$$g(L) = \frac{J(L)}{W(L)}, \quad \varphi(L) = \frac{\delta J(L)}{J(L)}, \qquad (14)$$

the latter having as an upper bound a certain value $\varphi_{\max}$ (Sec. 2). A phase diagram in coordinates $(g, \varphi)$ is shown in Fig. 3. At $\varphi = 0$, the boundary between localized and delocalized states is located at $g \sim 1$. An increase in $\varphi$ leads to greater disorder in the system, and the boundary $AB$ between the two phases displaces to higher $g$ and tends to infinity at some $\varphi_c$ (a curve like $AB'$ precludes localization due to the pure off-diagonal disorder, when the Thouless number is infinite).

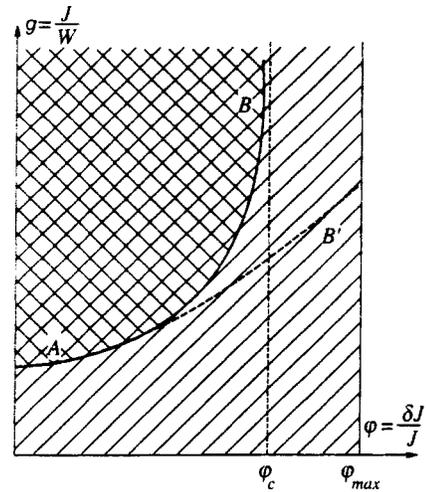

FIG. 3. Phase diagram in coordinates $(g, \varphi)$. The hatched area corresponds to localized states, the cross-hatched area to delocalized states.

The existence of the critical point $\varphi_c$ solves the problem of the second parameter in the renormalization group: the new nontrivial scale is associated not with $\varphi$, but with $\varphi - \varphi_c$.

If the parameters $g$ and $\varphi$ uniquely determine the state of a disordered system, then in the course of the Thouless scale transformation one point of plane $(g, \varphi)$ turns into another point of this plane. If the system is at a critical point, it can move only along the critical $AB$ surface, which is the locus of such points.

In order to return to the conventional scheme of one-parameter scaling, we should postulate, in accordance with the conventional concepts of the theory of critical phenomena (Ref. 17, Ch. 6), the existence of a fixed point $F$ (Fig. 4a), which is stable for states on the critical surface but unstable for states off the critical surface. In the theory of differential equations,[39] such a property is associated with a saddle point characterized by two asymptotes, $AB$ and $CD$, and hyperbolic trajectories in the vicinity of this point (Fig. 4a). Changes in the Thouless number $g(L)$ with scale $L$ for this case are shown in Fig. 5a. It has a constant value $g_c$ at point F (curve *1*), relaxes to $g_c$ at a finite scale $L_0$ for the points on the critical surface different from F (curves *2* and *3*), approaches $g_c$ at the scale $L_0$ and departs from this value at the scale $\xi$ for the points close to the critical surface (curves *4* and *5*). Roughly speaking, evolution in the $(g, \varphi)$ plane consists of two stages, namely the fast relaxation to the curve $CD$ and slow motion along this curve. At scales $L \gg L_0$ the $(g, \varphi)$ plane is in fact compressed to the line $CD$, and positions on this line are determined by the Thouless number.[2] Thus, we have returned to the conventional scheme, and we assume it to be valid for low dimensions.

Suppose that there is no stationary point on the critical surface at large $d$. Then a system at a critical point moves upward along curve $AB$ as $L$ increases (Fig. 4b). The downward motion is impossible because this means that off-diagonal disorder disappears asymptotically at large $L$ and contradicts the physical arguments of Sec. 2. The Thouless number $g(L)$ increases with $L$ at the transition point (curve *1* in Fig. 5b), in the metallic phase it increases faster,[1] $g(L)$ $\sim \sigma L^{d-2}$ (curve 2), and in the localized phase the curve exhibits reentrant behavior (curve 3).

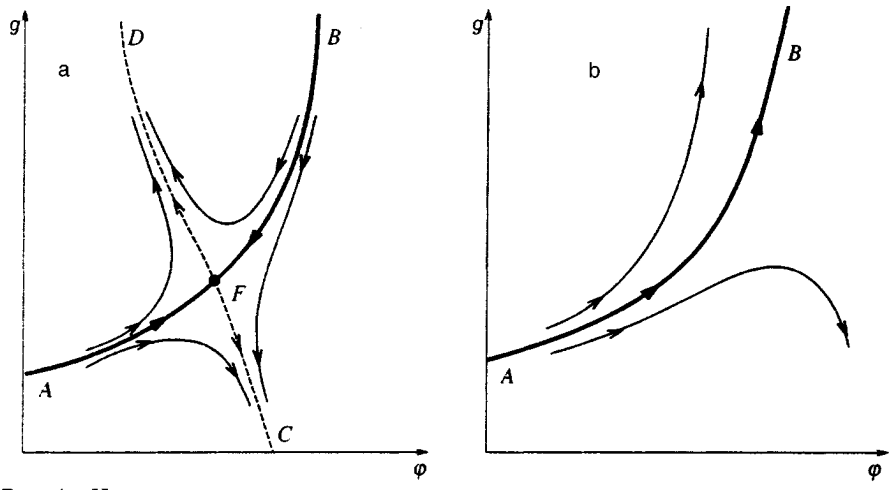

FIG. 4. Flow diagram for Thouless' scale transformations (a) in the presence of a stationary point F on the critical surface AB and (b) in the absence of such a point.

At first sight, such reentrant behavior is absurd from the physical standpoint. This means[1] that the degree of hybridization between block states increases at smaller $L$, but then drops for an unclear reason. In reality, this is not so, since the hybridization is not determined entirely by the Thouless number, but is also a function of $\varphi(L)$. At the transition point, the effective disorder (hence the hybridization degree) remains at the same level but is transferred from the diagonal type to off-diagonal one. In the localized phase, the effective disorder increases monotonically, but in the first stage the Thouless number grows, and the diagonal disorder characterized by this parameter decreases owing to transformation to the off-diagonal disorder. Only when $L > \xi$ and the total disorder has increased considerably does diagonal disorder also begin to grow.

As the space dimensionality increases, the first scenario (Fig. 4a) should gradually transfer to the second one (Fig. 4b), so the stationary point should move upwards along the curve $AB$ and go to infinity at a certain dimensionality $d_{c2}$. We identify this value with the upper critical dimensionality. The aim of subsequent analysis is to develop a phenomenological theory of this bifurcation.

The phenomenological description is possible because the functions $\beta(g,h)$ and $\gamma(g,h)$ in the two-parameter scaling equations (8) admit regular expansions. By virtue of Eq. (7), they describe a relation between two finite systems, whereas all singularities emerge in the thermodynamic limit.[17] This argument assumes, however, an adequate choice of scaling variables, which do not have their built-in singularities. In this sense, the variable $\varphi$ is not appropriate because it has a singular point $\varphi_c$. Therefore we introduce a new variable $h = F(g,\varphi)$ such that in the $(g,h)$ plane the curves of Fig. 4a take the form shown in Fig. 6, i.e., curve $AB$ has an asymptote $g \sim h$ as $g,h \to \infty$ and curve $CD$ becomes a vertical line. The first condition is adopted so that the critical surface, which is associated with no singularities, should have regular projections on both coordinate axes, and the second is assumed to simplify the equations (see below).

In investigating the bifurcation, it is sufficient to analyze Eq. (8) in the region of large $g$ and $h$, where it can be transformed to

$$\frac{d\ln g}{d\ln L} = (d-2) + \frac{Ah}{g} + \frac{Bh^2}{g^2} + \frac{Ch^3}{g^3}$$
$$+ \ldots \equiv (d-2) + \tilde{\beta}\left(\frac{g}{h}\right), \qquad (15a)$$

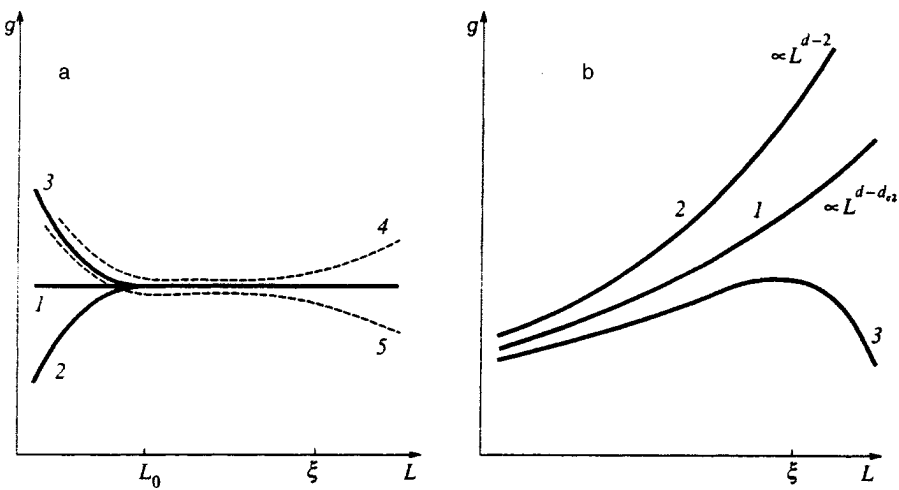

FIG. 5. Evolution of the Thouless parameter in scenarios illustrated by Figs. 4a and 4b.

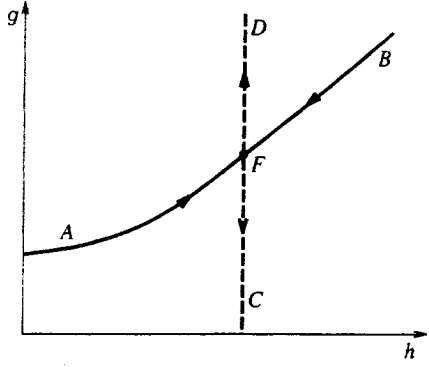

FIG. 6. Diagram of Fig. 4a after the variable change $h=F(g,\varphi)$.

$$\frac{d \ln h}{d \ln L} = \mu + \frac{b}{h}, \quad (15b)$$

where parameter $\mu$ changes sign at $d=d_{c2}$,

$$\mu = \alpha(d-d_{c2}), \quad d \to d_{c2}, \quad (16)$$

and $\alpha>0$, $b>0$, and $A<0$. Indeed, at $h=$const all conclusions from Ref. 1 apply to function $\beta(g,h)$, i.e., it has asymptotes $d-2$ and $\ln g$ for large and small $g$, and at $d>2$ has a root $g_c$, which is a function of $h$ in this specific case. By expanding $\beta(g,h)$ in powers of $1/g$,

$$\beta(g,h) = (d-2) + \frac{A_1(h)}{g} + \frac{A_2(h)}{g^2} + \ldots, \quad (17)$$

we find that the expansion $A_n(h)$ in powers of $1/h$ should begin with $h^n$ in order to yield a root $g_c \sim h$ (Fig. 6). By retaining the leading terms of the expansion in $h$, we obtain Eq. (15a).

As follows from the foregoing, at $d>d_{c2}$ the function $\gamma(g,h)$ should lead to unbounded growth in $h$, which, however, should not be faster than that in $g$, so that the root $g_c \sim h$ should retain its physical sense. Given that $g(L)$ increases no faster than $L^{d-2}$,[1] we have at large $h$ the condition $0<\gamma(g,h)<d-2$, which indicates that the expansion of $\gamma(g,h)$ in powers of $1/g$ and $1/h$ begins with a zero-order term:

$$\gamma(g,h) = \mu + \frac{a}{g} + \frac{b}{h} + \ldots \quad (18)$$

If the variables are defined so that curve CD is a vertical line, the coordinate $h_c$ of the stationary point is independent of $g$ and the coefficient $a$ in Eq. (18) is zero. The stationary point should be stable for $d<d_{c2}$, and absent for $d>d_{c2}$, which means that $b$ is positive and $\mu$ changes sign at $d=d_{c2}$, as can be seen in Eqs. (15) and (16).

Equation system (15) is easy to analyze. For $d<d_{c2}$, Eq. (15b) has a stationary point $h_c=b/|\mu|$, and the variable change $g \to gh_c$ in Eq. (15a) returns us to the one-parameter scaling with the critical exponents given by equations

$$1/\nu = g_c \tilde{\beta}'(g_c), \quad s = \nu(d-2), \quad (d-2) + \tilde{\beta}(g_c) = 0. \quad (19)$$

For $d>d_{c2}$ and large $h$, we have $h(L) \propto L^\mu$, and after the change $g \to gL^\mu$, Eq. (15b) is reduced to a one-parameter form, but with $d-2-\mu$ instead of $d-2$. For $L \lesssim \xi$, the Thouless number follows the law

$$g(L) = g_c \left(\frac{L}{a_0}\right)^\mu + (g_0-g_c)\left(\frac{L}{a_0}\right)^{\mu+1/\nu}, \quad (20)$$

and the critical exponents are determined by the equations

$$1/\nu = g_c \tilde{\beta}'(g_c), \quad (21a)$$

$$s = \nu(d-2-\mu), \quad (21b)$$

$$(d-2-\mu) + \tilde{\beta}(g_c) = 0. \quad (21c)$$

The localization radius is defined as the distance at which the parameter $g(L)$ begins to drop for $g_0<g_c$ (i.e., in the localized phase) and the exponent $s$ is determined by matching the function defined by Eq. (20) and $g(L) \sim \sigma L^{d-2}$ at $L \sim \xi$. At the transition point, the Thouless number increases according to the law

$$g(L) \propto L^\mu, \quad (22)$$

which is the reason why the Wegner relation fails (see Eq. (21b)). The comparison between Eqs. (19) and (21) demonstrates that critical exponents as functions of $d$ have cusps at $d=d_{c2}$.

Usually, one feature of the upper critical dimensionality is that the critical exponents are independent of $d$ above $d_{c2}$. As follows from Eq. (21b), this is possible if $\mu=d+$const, which yields in combination with Eq. (16)

$$\mu = d - d_{c2}. \quad (23)$$

Given this relation, we obtain the Thouless number as a function of the length scale for $L \lesssim \xi$:

$$g(L) = g_c + (g_0-g_c)(L/a_0)^{1/\nu}, \quad d<d_{c2}, \quad (24a)$$

$$g(L) = g_c(L/a_0)^{d-d_{c2}} + (g_0-g_c)(L/a_0)^{d-d_{c2}+1/\nu},$$

$$d>d_{c2}. \quad (24b)$$

Equation (24b) is the main result of our phenomenological approach. Equation (24a) is a well-known consequence of one-parameter scaling, but its range of applicability is limited.

## 5. COMPARISON TO THE SYMMETRY THEORY

The symmetry theory[30] yields the same values of critical exponents as the Vollhardt and Wölfle self-consistent theory[32]:

$$\nu = 1/(d-2), \quad s=1 \quad \text{for} \quad 2<d<4,$$
$$\nu = 1/2, \quad s=1 \quad \text{for} \quad d>4. \quad (25)$$

For $d<4$ they are compatible with the one-parameter scaling because the Wegner relation $s=\nu(d-2)$ holds. Its failure at $d>4$ means that $d_{c2}=4$.

In order to compare the results given by Eq. (24) to the symmetry theory, let us derive from the latter[30] the diffusion coefficient $D_L$ for a finite block of size $L$. It is calculated using the diffusion coefficient $D(\omega,q)$ for an infinite system using the formula[3)]

$$D_L \sim D\left(i\frac{D_L}{L^2}, L^{-1}\right). \quad (26)$$

It was shown in Ref. 30 that

$$D(\omega,q) = D_0(\omega) q^0, \quad q \ll a_0^{-1}, \quad (27)$$

and $D_0(\omega)$ is given by the equation

$$D_0(\omega) = A\tau + B\left(-\frac{i\omega}{D_0(\omega)}\right)^{1/2\nu}, \quad (28)$$

where $\tau$ is the distance to the transition point. Given that $g(L) \propto D_L L^{d-2}$ and parameter $\tau$ is proportional to $g_0 - g_c$, we can easily derive from Eqs. (26)–(28)

$$g(L) = g_c(L/a_0)^{d-2-1/\nu} + (g_0-g_c)(L/a_0)^{d-2}. \quad (29)$$

This result is similar to Eq. (24) but not identical in the general case. The results expressed by Eqs. (24) and (29) are identical only for specific values of critical exponents given by Eq. (25):

$$\begin{aligned} g(L) &= g_c + (g_0-g_c)(L/a_0)^{d-2}, \quad d<4, \\ g(L) &= g_c(L/a_0)^{d-4} + (g_0-g_c)(L/a_0)^{d-2}, \quad d>4. \end{aligned} \quad (30)$$

Thus, the phenomenological model developed in the reported work is in full agreement with the symmetry theory.[30] This correspondence between the two theories is far from trivial because the symmetry theory is based on different principles and does not use in any way the scaling concepts.

This work was stimulated by discussions with V. E. Kravtsov, A. D. Mirlin, and M. V. Feigel'man, who are gratefully acknowledged. I am also grateful to participants of seminars at the Kapitza Institute for Physical Problems and Lebedev Institute of Physics for interesting discussions on the results of this research.

This work was supported by INTAS (application No. 580) and Russian Fund for Fundamental Research (Project 96-02-19527).

*)E-mail: suslov@kapitza.ras.ru

1)"We cannot see how any statistical feature of the energy levels other than this … ratio can be relevant" (Ref. 1).

2)The assumption that only two parameters, $g$ and $\varphi$, are essential means in reality that all other parameters relax rapidly to a surface which can be mapped one-to-one onto the $(g,\varphi)$ plane.

3)Equation (26) fails in the localized phase for $L \gtrsim \xi$ due to a nonlocal response.[32]

Translation was provided by the Russian Editorial office.